\documentclass[10pt,a4paper]{article}
\usepackage{amsfonts,amsmath,amsthm,amssymb,mathtext,latexsym}
\usepackage[english]{babel}

\newtheorem{thm}{Theorem}[section]
\newtheorem{lem}[thm]{Lemma}
\theoremstyle{definition}
\theoremstyle{remark}
\numberwithin{equation}{section}

\begin{document}
\title{To the question of the existence and uniqueness of the scattering problem solution. The case of three
 one-dimensional quantum particles interacting by finite repulsive pair potentials.}
\author{A.M. Budylin, S.B. Levin}
\date{}
\maketitle
%\tableofcontents
%%%%%%%%%%%%%%%%%%%%%%%%%%%%%%%%%%%%%%%%%%%%%%%%%%%%%%%%%%%%%%%%%%%%

\section*{Introduction}
In the work \cite{BL1} asymptotic formulas at infinity in configuration space were offered for the first time
for absolutely continuous spectrum eigenfunctions uniform at angle variables. We considered the case of
three one-dimensional short-range quantum particles interacting by the repulsive pair potentials.
The mentioned asymptotic formulas
were obtained in terms of formal asymptotic decompositions
within the framework of a rather subtle heuristic analysis.

The present work aims at announcing a proof of the existence and uniqueness theorem
of the scattering problem solution in the finite repulsive pair potentials case. It is worth emphasizing that the limitation
of treatments for the finite potentials case does not lead
to a problem simplification in its essence as the interaction potential
of all three particles remains non decreasing at infinity
but allows us to switch off from a certain number of technical details.

\section{ Preliminaries}
 \paragraph{Initial Setting and Reducing of the Problem.}

In the initial setting the non-relativistic Hamiltonian $H$ is considered
\begin{equation*}
H\psi=-\Delta \psi +\frac{1}{2}\sum_{1\leqslant i\ne j\leqslant 3}v(z_{i}-z_{j})\psi\,,
\quad z_{i}\in \mathbb{R}\,,
\quad \boldsymbol{z}=(z_{1},z_{2},z_{3})\in \mathbb{R}^{3}\,,\quad
\psi=\psi(\boldsymbol{z})\in\mathbb{C}
\,,
\end{equation*}
$\Delta$ is a Laplace operator in  $\mathbb{R}^{3}$, $v$
is an even finite integrable function
$\mathbb{R}^{3}\to [0,+\infty)$, defining two-particle interaction.
In this case the essential self-adjointness of the operator $H$
in the square-integrable functions space is well known.

To separate the center-of-mass motion, we restrict the Hamiltonian
on the surface $\Pi$ which is denoted by the equation $\sum z_{i}=0$.
With a certain negligence, we will hereafter denote the Laplace-Beltrami operator
on the plane  $\Pi\subset \mathbb{R}^{3}$ by $\Delta$.
Here at $\Pi$
it is convenient to use any pair of $(x_{i},y_{i})\,,\; i=1,2,3,$, the so called
Jacobi coordinates, uniquely denoted by the equations
$x_{i}= \tfrac{1}{\sqrt{2}} (z_{k}-z_{j})\,, \; y_{i}=\sqrt{\tfrac{3}{2}}z_{i}\,,$
the indices $(i,j,k)$ here form even permutations.

In view of the orthonormality of the  Jacobi coordinates and the  Laplace-Beltrami operator invariance
we have $\Delta= \frac{\partial^{2}}{\partial x_{i}^{2}}+\frac{\partial^{2}}{\partial y_{i}^{2}}\,$
and our operator takes on the form
\begin{equation}
\label{osnovnoj-H}
H=-\Delta +V\,,\quad V=\sum_{i=1}^{3} v_{i}\,,\quad v_{i} (x_{i},y_{i}) =v(x_{i})\,.
\end{equation}
Note that the support of the potential $V$ lies in an infinite cross domain.

Define the resolvent of the operator $H$:
$\;R (\lambda)= (H-\lambda I)^{-1}\,,\quad \lambda\notin[0,+\infty)\,.$
Here and hereafter $I$ is an identity operator. The resolvent  $(H_{0}-\lambda I)^{-1}$
of the unperturbed operator $H_{0}=-\Delta$ will be denoted by $R_{0} (\lambda)$.

 \paragraph{Saturable Absorption Principle.}
We consider the scattering problem within the framework of a so called
stationary approach. In this case a wave operators treatment is replaced by
the study of limit values $R(E\pm i0)$ of the considered resolvent,
when a spectral parameter $\lambda$ approaches the real axis
($\lambda\rightsquigarrow E\pm i0\,,\; E\in (0,+\infty)$).

The proof of the existence of such limit values in the appropriate topology
is the very essence of the saturable absorption principle. Once the existence
of the resolvent limit values is determined, the study of the
wave operators takes the known standard form, see, for example,
 ~\cite{Yaf},\cite{RS3}.

The existence of the resolvent limit values $R (E\pm i0)\,$
is most often considered in the following weak sense (a so called
rigged Hilbert space method, see ~\cite{GelVil}). In this case
a certain Banach space $\mathcal{B}$ is continuously include
into the main Hilbert space $\mathcal{H}$, and this, in its turn, allows to include
$\mathcal{H}$ into the conjugate to $\mathcal{B}$ space
$\mathcal{B}^{*}$ with a further proof that
$R (E\pm i\varepsilon): \; \mathcal{B}\to \mathcal{B}^{*}$
has a continuous extension at $\varepsilon \downarrow 0$.

Thus the generalized eigenfunctions in this case are treated as
the elements of the space $\mathcal{B}^{*}$, while the main object
of study becomes a scalar product
$(R (E\pm i\varepsilon)\varphi,\varphi)\,,\; \varphi\in \mathcal{B}\,,$
with $\varepsilon \downarrow 0$.

Hereafter, to be definite, we restrict ourselves to a consideration of the case
$\mathrm{Im}\lambda \downarrow 0$.

 \paragraph{ Friedrichs - Faddeev Model}

In the Friedrichs - Faddeev model,
see ~\cite{Fad},\cite{Yaf},
within the framework of the saturable absorption principle, an unperturbed operator
is treated as an operator of multiplication by an argument, while the perturbation
is an integral operator with a smooth kernel.
In the applications it essentially means a transformation into a dual
momentum representation. We support this point of view too, though
it seems natural to make a certain substantial analysis in an initial
coordinate representation.

For the Friedrichs - Faddeev model it is natural to choose as an auxiliary
space $\mathcal{B}$ the space of Holder functions. Our choice of an auxiliary space
will be also defined by this circumstance. In the topology of the auxiliary space
the limit in a weak sense values of the resolvent will be considered.

Such a space in a momentum representation will be the space
$H^{\mu,\theta} (\mathbb{R}^{2})$ $(0<\theta,\mu<1)$ of the Holder functions with a norm
\begin{equation}
\label{geld-norma}
\|f \|_{\mu,\theta}=\sup\limits_{\xi,\eta} (1+|\xi|^{1+\theta})\Bigl( |f(\xi)|+
\frac{|f (\xi+\eta)-f (\xi)|}{
|\eta|^{\mu}}\Bigr)\,.
\end{equation}

The space of the functions in the coordinate representation, the Fourier images of which
lie in $H^{\mu,\theta} (\mathbb{R}^{2})$ we will denote by
$\hat{H}^{\mu,\theta}$.

The analysis of resolvent singularities in a momentum representation
can be conveniently made within the framework of a so called alternating Schwartz method,
see ~\cite{Mor},\cite{BB1},\cite{BBsin}.
Note that the known Faddeev equations, see ~\cite{Fad},
can be also interpreted as a certain version of the alternating method.

\paragraph{ Alternating Schwartz Method.}

In the problem considered there stands out first of all a possibility of separating of variables
for a partial Hamiltonian
$\label{chastichnyj-H}
H_{i}=-\Delta+v_{i}\,.$

Thus, the existence of the limit values of resolvent
$R_{i} (\lambda)=
(H_{i}-\lambda I)^{-1}$ is easily controlled. As a consequence, there arises a question
about accounting such contributions into resolvent $R(\lambda)$
of a complete Hamiltonian $H$
with a total potential $V=\sum v_{i}$.

The scheme of such accounting is known in literature under the name of
the alternating Schwartz method.
Denote by $\{G_{i}\}|_{i=1}^{n}$ a certain set of linear operators
in a complex vector space $\mathcal X$. Define the operator
$G=\sum_{i=1}^n G_i\,.$
Assuming that all operators $I-G_i$  are bijective, let
$\label{gamma-i}
I-\Gamma_{i}=(I-G_i)^{-1}\,.$ Operator
$\Gamma_{i}$
is called an inversion operator relative to the operator $G_{i}$.

The essence of the alternating Schwartz method resolves itself
into the following.
Bijectivity of the operator matrix
$
\boldsymbol{L}=\begin{pmatrix}I&\Gamma_1&\ldots&\Gamma_1\\
                             \Gamma_2&I&\ldots&\Gamma_2\\
                             \vdots&\vdots&\ddots&\vdots\\
                             \Gamma_n&\Gamma_n&\ldots&I
\end{pmatrix}
$
in space ${\mathcal X}^n$ is equivalent to the bijectivity of the operator
$I-G$ in space $\mathcal X$. Moreover, if the operator matrix
$\boldsymbol{\gamma}$ is a solution of the equation
$
\boldsymbol{L}\cdot\boldsymbol\gamma=\mathrm{diag}(\Gamma_1,\ldots\Gamma_n)\,,
$
where by $\mathrm{diag} (\Gamma_{1},\ldots \Gamma_{n})$ a diagonal matrix is denoted,
then the operator $\Gamma=\sum
\gamma_{ij}\,,$ where summation extends to all matrix elements of operator matrix
$\boldsymbol{\gamma}= (\gamma_{ij})$, defines an inverse to $I-G$ operator by
the equality $(I-G)^{-1}=I-\Gamma$.

Note that if the operator matrix $\boldsymbol{\Gamma}=\boldsymbol{L}-I$
in an adequate Banach space is restricted and its norm is less than unity, then
\begin{equation*}
\Gamma=\sum \Gamma_{i}-\sum_{i\ne j} \Gamma_{i}\Gamma_{j}+
\sum_{i\ne j\ne k} \Gamma_{i}\Gamma_{j}\Gamma_{k}-\ldots
\end{equation*}
where a series of sums is taken as convergent in norm. Exactly this formula  explains
the name of the method.

Finally, note that the equality
$(I+\boldsymbol{\Gamma})^{-1}=
(I-\boldsymbol{\Gamma}^2)^{-1}(I-\boldsymbol{\Gamma})\,,$
reduces the inversion of the operator matrix  $\boldsymbol{L}$
to the inversion of the operator matrix $I-\boldsymbol{\Gamma}^{2}$.

But the matrix $\boldsymbol{\Gamma}^{2}$
as its components has the sums of the operators as
$\Gamma_{i}\Gamma_{j}$ at $i\ne j$.

Applied to the considered problem it is the analysis of such products $\Gamma_{i}\Gamma_{j}$ that
has brought the desired result.
To embed our problem into this scheme all we need is to separate a free resolvent
$R_{0} (\lambda)$.
As this takes place
$
R (\lambda)=R_{0} (\lambda) (I-G (\lambda))^{-1}\,,\;
G (\lambda) =\sum G_{i} (\lambda)\,,\;
G_{i} (\lambda) =v_{i}R_{0} (\lambda)\,.
$
The reflection operators with respect to
$G (\lambda)$ or $G_{i} (\lambda)$
will be denoted, accordingly, by
$\Gamma (\lambda)$ и $\Gamma_{i} (\lambda)$.
Thus,
\begin{equation}
  \label{eq:1}
  R (\lambda)=R_{0} (\lambda) (I-\Gamma (\lambda))\,.
\end{equation}

\section{Short Summary of the Results }

Further we assume that
\textit{the Fourier transform of the finite function $v$, defining a two-particle interaction,
belongs to
$H^{\mu_{0},\theta_{0}} (\mathbb{R})$ at certain $\mu_{0}>0$ и $\theta_{0}>0$}.

The integral kernel of the operator $\Gamma_{j} (\lambda)$ looks as
$
\Gamma_{j} ( \boldsymbol{z}, \boldsymbol{z}'|\lambda)=
v(x_{j}) \iint\limits_{\mathbb{R}^{2}} d \boldsymbol{q} \,\frac{\psi_{j} (\boldsymbol{z},
\boldsymbol{q})\overline{\psi_{j} (\boldsymbol{z}', \boldsymbol{q})}}
{\boldsymbol{q}^{2}-\lambda}\,,
$
here $d \boldsymbol{q}$ is a Lebesgue measure at the momentum variable plane
$\boldsymbol{q}= (k,p)$,
dual to $\boldsymbol{z}= (x_{j},y_{j})$ and $\boldsymbol{z}'=
(x'_{j},y'_{j})$, while а $\psi_{j}$
is a continuous spectrum generalized eigenfunction of the operator
$H_{j}$. It looks as
$
\psi_{j}(\boldsymbol{z},\boldsymbol{q})=\varphi_{j}(x_{j},k)e^{iy_{j}p}\,,
$
where
$\varphi_{j}(x,k)$ is a solution of a one-dimensional Schroedinger equation
$
\bigl(-\frac{d^2}{dx^2}+v(x)\bigr)\varphi(x,k)=k^2\varphi(x,k)
$
and is defined by the scattering data of a corresponding one-dimensional scattering
problem. In a certain sense a key to the proof of the announced result is
the computation of the asymptotic of the kernel
$\Gamma_{j} (\lambda)\Gamma_{k} (\lambda)$
with $|\boldsymbol{z}|\to\infty$.
It permitted to obtain a partition of the operator
$\Gamma_{j} (\lambda)\Gamma_{k} (\lambda)$ into a sum
$
\Gamma_{j} (\lambda)\Gamma_{k} (\lambda) =A_{jk} (\lambda) + B_{jk} (\lambda)\,,
$
where $A_{jk} (\lambda)$ is the 2nd rank operator including all the
\flqq bad \frqq\ part of this product,i.e. the one which goes beyond the space
$L_{2} (\mathbb{R}^{2})$
with $\mathrm{Im}\lambda\downarrow 0$, while the operator $B_{jk} (\lambda)$
is a compact operator in $\hat H^{\mu,\theta}$, strongly continuous in $\lambda$
with $\mathrm{Im}\lambda\geqslant 0$ and
$0<c_{1}\leqslant \mathrm{Re}\lambda \leqslant c_{2}<\infty$,
if $\mu$ and $\theta$ are small enough.
 The limit operator $A_{jk} (E+i0)$ act on functions from $\hat{H}^{\mu,\theta}$
 into a two-dimensional space of the functions of the type
 $ v(x_{j})\varphi_{j}(x_{j},0)|y_{j}|^{-1/2}e^{i|y_{j}|\sqrt{E}}(C_{1}\chi (y_{j})+
C_{2} \chi (-y_{j}))$,
where $\chi$ can be defined as a smoothed characteristic function of the semi-axis
$(T,+\infty)$ with $T\gg 1$.

The space of such functions has square root analytical singularities on the axis
in a momentum representation, exactly - the singularities of the type
$(p\pm \sqrt{E})^{-1/2}$.
We will denote the described above range of the operator
$A_{jk} (\lambda)$ by the space of functions of the type $A_{j}$.

It is worth emphasizing that it is the necessity of an extraction of such
an operator as  $A_{jk}$ that is a distinguishing feature of this problem -
the three one-dimensional particles scattering problem - in comparison with
the case of the three three-dimensional particles scattering problem considered
in the work  ~\cite{Fad}.

The consequence of the partition  $\Gamma_{j} (\lambda)\Gamma_{k} (\lambda)$
is the representation
$
I- \boldsymbol{\Gamma}^{2} (\lambda) =I- \boldsymbol{A} (\lambda) -
\boldsymbol{B} (\lambda)\,,
$
where the operator matrices $\boldsymbol{A} (\lambda)$ and  $\boldsymbol{B} (\lambda)$
inherit the properties of the corresponding scalar operators
$A_{jk} (\lambda)$ and  $B_{jk} (\lambda)$.
The following assertion is valid.
\begin{lem}
  With $\mathrm{Im}\lambda>0$ the operator $(I- \boldsymbol{\Gamma}^{2})^{-1}$
allow a representation $(I- \boldsymbol{\Gamma}^{2})^{-1}=I-\tilde{\boldsymbol{A}} (\lambda)
-\tilde{\boldsymbol{B}} (\lambda)$ where the matrix components
$\tilde{\boldsymbol{A}} (\lambda)$ are the finite rank operators
acting into the algebraic sum of the spaces of the type $A_{j}\,,j=1,2,3\,$
while the matrix components $\tilde{\boldsymbol{B}}$
are compact operators in $\hat H^{\mu,\theta}$
strongly continuous in $\lambda$
with $\mathrm{Im}\lambda\geqslant 0$ and
$0<c_{1}\leqslant \mathrm{Re}\lambda \leqslant c_{2}<\infty$,
if $\mu$ and  $\theta$ are small enough.
\end{lem}
It is easy to see if $\varphi_{\lambda}$ is a function of
the type $A_{j}$ and $\psi\in \hat{H}^{\mu,\theta}$ then $(R_{0} (\lambda)\varphi_{\lambda},\psi)$
has a limit at $\mathrm{Im}\lambda\downarrow 0$.
As a consequence for the desired scalar operator $\Gamma (\lambda)$
the constructions of the alternating Schwartz method described above lead to the representation:
$
\Gamma (\lambda) = \sum \Gamma_{i} (\lambda) +N (\lambda)\,
$
where the operator $N (\lambda)$ possess the property that the product
$R_{0} (\lambda)N (\lambda)$ has a weak limit in $\hat H^{\mu,\theta}$
with $\mathrm{Im}\lambda\downarrow 0$ and
$0<c_{1}\leqslant \mathrm{Re}\lambda \leqslant c_{2}<\infty$,
if $\mu$ and  $\theta$ are small enough.
It means, see ~\eqref{eq:1} that the following is valid
\begin{thm}
   $R (\lambda)$ has a weak limit in $\hat H^{\mu,\theta}$
with $\mathrm{Im}\lambda\downarrow 0$ and
$0<c_{1}\leqslant \mathrm{Re}\lambda \leqslant c_{2}<\infty$,
if $\mu$ and $\theta$ are small enough.
\end{thm}

Note that this theorem guarantees a validity of the asymptotic constructions of the works
~\cite{BL1},\cite{BLNO}.

\end{document}